\documentclass[final]{IEEEtran}
\usepackage{cite}
\usepackage{graphicx}
\usepackage{picinpar}
\usepackage[cmex10]{amsmath}
\usepackage{amsmath,amsfonts,amssymb}
\usepackage{subfigure}

\usepackage{algorithm}
\usepackage{algorithmic}
\usepackage{stfloats}
\usepackage{bm}
\usepackage{color}
\usepackage{stfloats}
\usepackage{url}
\usepackage{makecell}
\usepackage{multirow}
\usepackage{array}
\usepackage{booktabs}

\begin{document}
\pdfoptionpdfminorversion=6
\newtheorem{lemma}{Lemma}
\newtheorem{corol}{Corollary}
\newtheorem{theorem}{Theorem}
\newtheorem{proposition}{Proposition}
\newtheorem{definition}{Definition}
\newcommand{\e}{\begin{equation}}
\newcommand{\ee}{\end{equation}}
\newcommand{\eqn}{\begin{eqnarray}}
\newcommand{\eeqn}{\end{eqnarray}}
\renewcommand{\algorithmicrequire}{ \textbf{Input:}} 
\renewcommand{\algorithmicensure}{ \textbf{Output:}} 

\title{Data-Driven Deep Learning to Design Pilot and Channel Estimator For Massive MIMO}

\author{Xisuo Ma, Zhen Gao,~\IEEEmembership{Member,~IEEE} ~\IEEEmembership{} ~\IEEEmembership{}\\
~\IEEEmembership{}
\thanks{X. Ma and Z. Gao are with both Advanced Research Institute of Multidisciplinary Science (ARIMS) and School of Information and Electronics,
Beijing Institute of Technology (BIT), Beijing 100081, China (E-mail: gaozhen16@bit.edu.cn).}
}
\maketitle

\begin{abstract}
In this paper, we propose a data-driven deep learning (DL) approach to jointly design the pilot signals and channel estimator for wideband massive multiple-input multiple-output (MIMO) systems.
By exploiting the angular-domain compressibility of massive MIMO channels, the conceived DL framework can reliably reconstruct the high-dimensional channels from the under-determined measurements.
Specifically, we design an end-to-end deep neural network (DNN) architecture composed of dimensionality reduction network and reconstruction network to respectively mimic the pilot signals and channel estimator, which can be acquired by data-driven deep learning.
For the dimensionality reduction network, we design a fully-connected layer by compressing the high-dimensional massive MIMO channel vector as input to low-dimensional received measurements, where the weights are regarded as the pilot signals.
For the reconstruction network, we design a fully-connected layer followed by multiple cascaded convolutional layers, which will reconstruct the high-dimensional channel as the output.
By defining the mean square error between input and output as loss function, we leverage Adam algorithm to train the end-to-end DNN aforementioned with extensive channel samples.
In this way, both the pilot signals and channel estimator can be simultaneously obtained.
The simulation results demonstrate that the superiority of the proposed solution over state-of-the-art compressive sensing approaches.
\end{abstract}
\vspace*{-1.5mm}
\vspace*{-2.0mm}
\IEEEpeerreviewmaketitle

\vspace{-1.5mm}
\section{Introduction}
Massive multiple-input multiple-output (MIMO) has been widely recognized as a promising technology in future wireless communications due to its high spectrum efficiency and large beamforming gain \cite{angle_sparse}.
Moreover, to mitigate the hardware cost and power consumptions imposed by fully-digital massive MIMO, the hybrid MIMO architecture with much reduced number of radio frequency (RF) chains has been proposed by dividing the beamforming into the cascaded analog domain and digital domain operations \cite{DL_precoding,precoding,mao,PAPR}.

For massive MIMO systems, the accurate channel state information (CSI) is essential for beamforming and signal detection.
However, it is challenging to accurately estimate the high-dimensional massive MIMO channels using low training overhead \cite{WC}. For frequency division duplexing (FDD) massive MIMO, since the downlink/uplink channel reciprocity does not exist, the downlink high-dimensional massive MIMO channel has to be estimated at the users, which are usually equipped with very few antennas.
To this end, by exploiting the compressibility of massive MIMO channels represented in the angular domain and/or delay domain, several low training overhead channel estimation solutions have been proposed \cite{angle_sparse,PAPR,WC,SBL,WZW}.
Specifically, in \cite{angle_sparse}, a spatially common sparsity based adaptive channel estimation and feedback scheme for FDD massive MIMO was proposed to reliably estimate and feed back the downlink CSI with significantly reduced overhead.
In \cite{PAPR}, A. Liao et al. utilized ESPRIT-type algorithms to acquire the super-resolution estimates of the angle of arrivals/departures (AoAs/AoDs) with high accuracy for millimeter-wave (mmWave) massive MIMO with hybrid beamforming.
Additionally, in \cite{SBL}, the authors developed an expectation maximization (EM)-based sparse Bayesian learning (SBL) channel estimation scheme for the time-varying massive MIMO systems.
In \cite{WZW}, Z. Wan et al. proposed a compressive sensing (CS)-based channel estimation solution for full-dimensional (FD) lens-array.
However, for sparse channel estimation schemes based on CS or spatial spectrum estimation techniques, due to the high-dimensional channel matrices to be estimated in massive MIMO systems, the involved matrix inversion operations can pose the prohibitively high computational complexity and storage requirement.

Recently, deep learning (DL) has shown its great potential to revolutionize communication systems by applying deep neural network (DNN) to various communication and signal processing problems \cite{bm1,Qin_WC,DL_for_5G}, which include modulation recognition \cite{modulationrecognition,modulationrecognition2}, signal detection \cite{signaldetection}, CSI feedback \cite{csifeedback}, and channel estimation \cite{DL_DOA, channelestimation, noFD}, network routing and traffic control \cite{bm2,bm3,bm4}, et al.
Specifically, in \cite{csifeedback}, a novel CSI sensing and recovery mechanism, called CsiNet, was developed to recover CSI with improved reconstruction quality and reduced feedback overhead, which was closely related to the autoencoder in DL.
In \cite{DL_DOA}, a DL-based channel estimation and direction-of-arrival (DOA) estimation solution was proposed for massive MIMO systems, where the DNN was exploited to learn the statistical characteristics of wireless channels and the spatial structure in the angle domain.
In \cite{channelestimation}, by unfolding the iterations of denoising based approximate message passing (DAMP) and utilizing denoising neural network in each iteration, a learned DAMP (L-DAMP) algorithm was proposed to estimate the beamspace channel.
In \cite{noFD}, P. Dong et al. proposed a deep convolutional neural network (CNN)-based channel estimation solution, where the temporal correlation of time-varying channels was exploited.

Nevertheless, for most of existing DL-based channel estimation approaches, the pilot design at the transmitter and the channel estimation at the receiver are separately designed, which may lead to some performance loss.
In this work, inspired by the idea of autoencoder in \cite{DL_for_5G}, we propose an end-to-end DNN architecture to jointly design the pilot signals (encoding) and channel estimator (decoding).
Specifically, for the dimensionality reduction network, we design a fully-connected layer by compressing the high-dimensional channel vector as input to a low-dimensional vector as the received measurements, where the weights are regarded as the pilot signals.
Then, for the reconstruction network, we design a channel estimator composed of a fully-connected layer and multiple cascaded convolutional layers to reconstruct the high-dimensional channel as the output.
By defining the mean square error (MSE) between the input and output as loss function, we adopt adaptive moment estimation (Adam) algorithm to train the end-to-end DNN architecture, so that the pilot signals and channel estimator can be jointly trained.
The simulation results demonstrate that the superiority of the proposed solution in comparison to state-of-the-art CS approaches.

The notations used in the paper and their definitions are listed in Table \ref{table} as follows.

\begin{table}[ht]
  \renewcommand\arraystretch{2.0}
  \centering
  \caption{Notations description}\label{table}
  \begin{tabular}{c|c}
  \Xhline{1.2pt}
  {\bf{Notation}} & {\bf{Definition}} \\
  \Xhline{1.2pt}
  Boldface lower case & Column vector \\
  \hline
  Boldface upper case & Matrix \\
  \hline
  $(\cdot)^{T}$ & Transposition operation \\
  \hline
  $(\cdot)^{H}$ & Conjugate transposition operation \\
  \hline
  ${\| \cdot \|}_{F}$ & Frobenius norm \\
  \hline
  ${\otimes}$ & Kronecker product \\
  \hline
  $\rm{Re} \{ \cdot \}$ & Real part  \\
  \hline
  $\rm{Im} \{ \cdot \}$ & Imaginary part \\
  \Xhline{1.2pt}
\end{tabular}
\end{table}

\vspace{-5mm}
\section{System Model}
\begin{figure}[t]
\centering
{\subfigure[]{ \label{fig1:cluster(a)}
\centering
\includegraphics[scale=0.65]{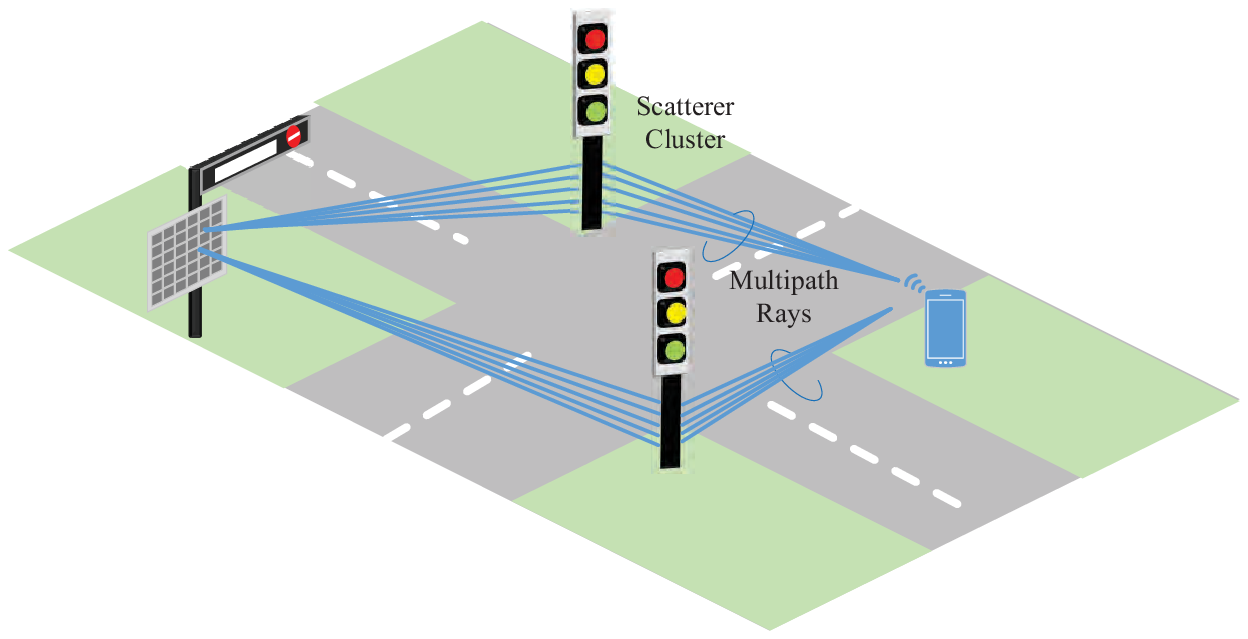}
}}
{\subfigure[]{ \label{fig1:onering(b)}
\centering
\includegraphics[scale=0.42]{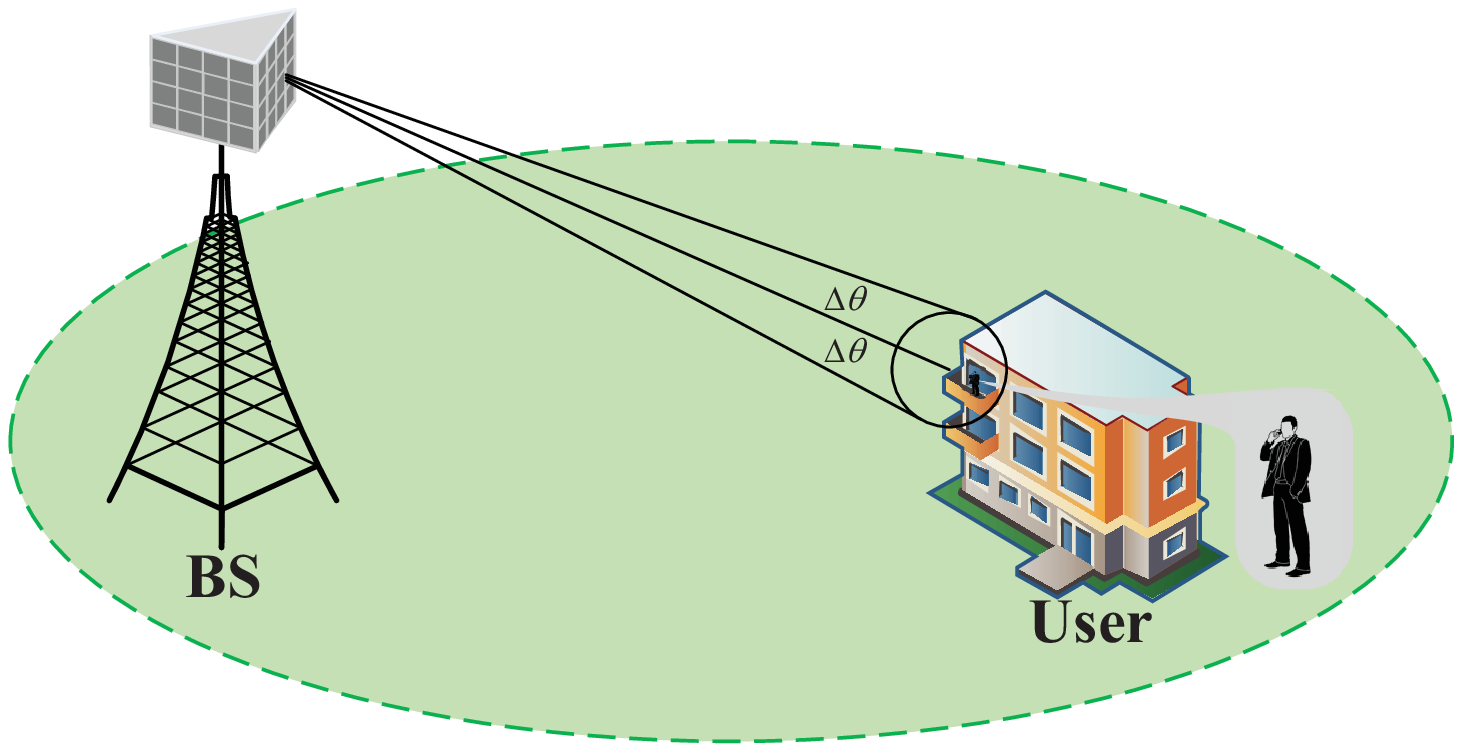}
}}
\caption{Channel scenarios considered in this work: (a) Cluster sparse channel model for mmWave massive MIMO systems, and (b) one-ring channel model for low-frequency massive MIMO systems.}
\label{fig1:channel_model}
\end{figure}

We firstly consider the downlink channel estimation for massive MIMO systems$\footnote{The proposed scheme aims to use the much reduced pilot overhead to estimate the high-dimensional massive MIMO channels, even this is an under-determined problem. However, for conventional small-scale MIMO systems, this challenge does not appear, and the classical least square (LS) estimation can achieve good performance.}$ with hybrid beamforming as shown in Fig. \ref{fig1:channel_model}. Here the massive MIMO systems can work at low-frequency band or mmWave frequency band, and orthogonal frequency division multiplexing (OFDM) with $K$ subcarriers is adopted.
The BS deploys $N_{\rm{BS}} = N_h \times N_v$ antennas in the form of uniform planar array (UPA) and $N_{\rm{RF}}$ RF chains to support $U$ single-antenna users, where $ N_h $ and $ N_v $ are the numbers of antennas in the horizontal and vertical directions, respectively.
In the downlink channel estimation phase, to estimate the $k$-th subcarrier's channel, the signal received at the $u$-th user in the $t$-th time slot (here one OFDM symbol is a time slot) can be expressed as
\begin{equation}\label{received2:signal}
{y}_k^{t} = {\bf{h}}_k^{T} {\bf{V}}_{\rm{RF}}^{t} {\bf{v}}_{{\rm{BB}},k}^{t} + {n}_k^t,
\end{equation}
where the user index $u$ is dropped to simplify the notations, ${\bf{h}}_k \in \mathbb{C}^{N_{\rm{BS}} \times 1}$ is the $k$-th subcarrier's channel,
$ {\bf{V}}_{\rm{RF}}^{t} \in \mathbb{C}^{N_{\rm{BS}} \times N_{\rm{RF}}}$ and $ {\bf{v}}_{{\rm{BB}},k}^{t} \in \mathbb{C}^{N_{\rm{RF}} \times 1} $ respectively denote the analog and digital precoders in the channel estimation phase,
and $ {n}_k^t $ is complex additive white Gaussian noise (AWGN).
So the received signals
in $M$ successive time slots can be collected as
\begin{equation}\label{all_time1:received_signal}
{\bf{y}}_k^{T} = {\bf{h}}_k^{T} {\bf{V}}_{\rm{RF}} {\bf{V}}_{{\rm{BB}},k} + {\bf{n}}_k^{T},
\end{equation}
where ${\bf{y}}_k^{T} = \left[ {y}_k^{1}, {y}_k^{2}, \cdots, {y}_k^{M} \right] \in \mathbb{C}^{1 \times M}$,
$ {\bf{V}}_{\rm{RF}} {\bf{V}}_{{\rm{BB}},k} = \left[ {\bf{V}}_{\rm{RF}}^{1} {\bf{v}}_{{\rm{BB}},k}^{1}, {\bf{V}}_{\rm{RF}}^{2} {\bf{v}}_{{\rm{BB}},k}^{2}, \cdots, {\bf{V}}_{\rm{RF}}^{M} {\bf{v}}_{{\rm{BB}},k}^{M} \right] \in \mathbb{C}^{N_{\rm{BS}} \times M}$
and $ {\bf{n}}_k^{T} = \left[ {n}_k^1, {n}_k^2, \cdots, {n}_k^M \right] \in \mathbb{C}^{1 \times M}$.
For both mmWave and low-frequency bands, the $k$-th subcarrier's channel in (\ref{all_time1:received_signal}) between the BS and the user can be written as a uniform spatial channel model, i.e.,
\begin{equation}\label{channel:delay}
{\bf{h}}_k = \sqrt{\frac{N_{\rm{BS}}}{N_c N_p}} {\sum_{i=1}^{N_c}} {\sum_{l=1}^{N_p}} {\alpha}_{i,l} e^{-j2\pi \tau_{i,l} f_s \frac{k}{K}} {\bf{a}}_{\rm{BS}}(\theta_{i,l},\varphi_{i,l}),
\end{equation}
where $ \alpha_{i,l} \sim \mathcal{C} \mathcal{N} (0, 1)$ and $\tau_{i,l}$ denote the propagation gain and delay corresponding to the $l$-th path in the $i$-th cluster, respectively,
$ f_s $ is the system sampling rate, $ \theta_{i,l} $ and $ \varphi_{i,l} $ are the azimuth and elevation angles of departure (AoD) of the $l$-th path in the $i$-th cluster at the BS, respectively.
Since the UPA is equipped at the BS, the corresponding response vector ${\bf{a}}_{\rm{BS}} \in \mathbb{C}^{N_{\rm{BS}} \times 1}$ can be expressed as
\begin{align}
{\bf{a}}_{\rm{BS}}(\theta, \varphi) &= \frac{1}{\sqrt{N_{\rm{BS}}}}[1, \cdots, e^{-j\frac{2\pi d}{\lambda}(m \sin(\theta) \cos(\varphi) + n \sin(\varphi))},\notag\\
&\cdots, e^{-j\frac{2\pi d}{\lambda}((N_h-1) \sin(\theta) \cos(\varphi) + (N_v-1) \sin(\varphi))}]^T,
\end{align}
where $ 1 \le m \le N_h $ and $ 1 \le n \le N_v $ are the horizontal and vertical indices of the corresponding antenna elements, respectively, $ \lambda $ is the signal wavelength, and $ d $ is the distance between adjacent antenna elements, which usually equals half of the wavelength, i.e, $ d / \lambda$ = $ \frac{1}{2}$.
Specifically, for the mmWave channel model, we consider $N_c$ clusters with each containing $N_p$ paths. The central angle in each cluster follows the uniform distribution $\left[ -\frac{\pi}{3}, \frac{\pi}{3} \right]$, and the angle spread is $ \Delta \theta $.
For the low-frequency massive MIMO channel, the cluster sparse channel reduces to the one-ring channel model \cite{angle_sparse}, where $N_c$ = 1 and this unique cluster contains $N_p$ paths. The distribution of central angle in low-frequency scenario is the same as that of the mmWave spatial channel model, and a corresponding angle spread $\Delta \theta$ for low-frequency bands is also considered.
\begin{figure*}[t]
  \centering
  \includegraphics[scale=0.4]{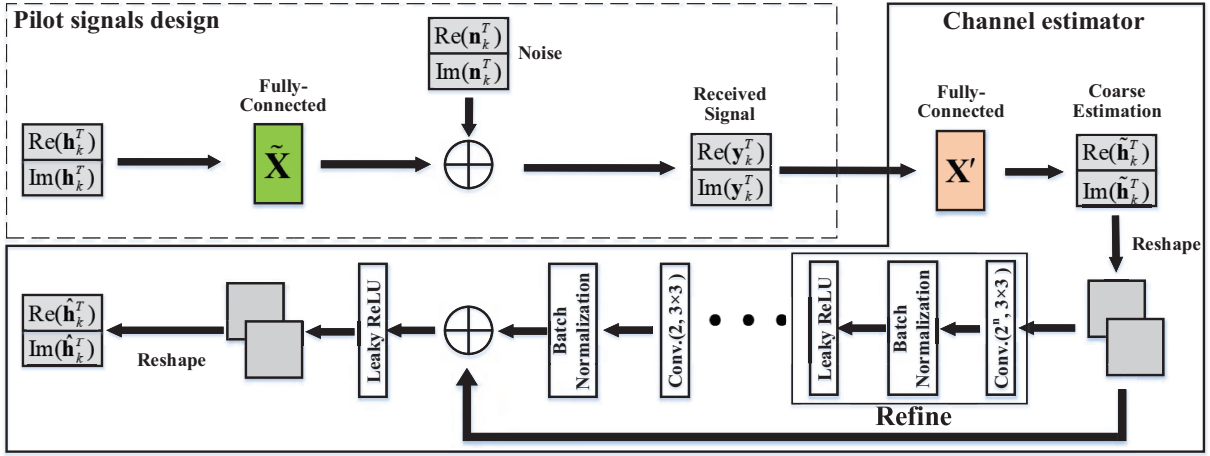}
  \caption{The proposed end-to-end DNN architecture can be used to jointly design the pilot signals and channel estimator.}
  \label{fig2:NN}
\end{figure*}
\vspace{-3mm}
\section{DNN-Based Channel Estimation Scheme}
Without loss of generality, we assume a simplified case$\footnote{This may not be practical in OFDM systems because the same signals for different subcarriers will lead to high peak to average power ratio (PAPR). Fortunately, according to \cite{PAPR}, we can introduce a pair of pseudo-random transmit scrambling and receive descrambling code over different pilot subcarriers to relax the assumption so that the high PAPR can be avoided effectively.}$ that ${\bf{V}}_{{\rm{BB}},k} = {\bf{V}}_{\rm{BB}}$ for $1 \le k \le K$.
The received signal in (\ref{all_time1:received_signal}) can be given by
\begin{equation}\label{new:received signal}
{\bf{y}}_k^{T} = {\bf{h}}_k^{T} {\bf{V}}_{\rm{RF}} {\bf{V}}_{\rm{BB}} + {\bf{n}}_k^{T}.
\end{equation}
Note that the channel ${\bf{h}}_k$ is generated according to the channel model aforesaid depending on the specific frequency band and scenarios.
In addition, according to \cite{precoding}, if the number of RF chains ${N_{\rm{RF}}}$ is greater than or equal to twice the total number of data streams ${N_s}$, the hybrid beamforming structure can realize any fully-digital beamformer exactly, regardless of the number of antenna elements. Thus for the case that $N_{\rm{RF}} \ge 2$, a solution to ${\bf{V}}_{\rm{RF}} {\bf{V}}_{\rm{BB}} = {\bf{V}}_{\rm{FD}}$ can be readily found.
By exploiting such a conclusion, (\ref{new:received signal}) can be given by
\begin{equation}\label{fd:received signal}
{\bf{y}}_k^{T} = {\bf{h}}_k^{T} {\bf{V}}_{\rm{FD}} + {\bf{n}}_k^{T} = {\bf{h}}_k^{T} {\bf{X}} + {\bf{n}}_k^{T},
\end{equation}
where ${\bf{X}} \in \mathbb{C}^{N_{\rm{BS}} \times M}$ denotes the transmitted signal after precoding.
By stacking all the subcarriers, the received signal $ {\bf{Y}} \in \mathbb{C}^{K \times M} $ can be written as
\begin{equation}\label{frequency spatial}
{\bf{Y}} = {\bf{H}}_s {\bf{X}} + {\bf{N}},
\end{equation}
where $ {\bf{Y}} = \left[ {\bf{y}}_1, {\bf{y}}_2, \cdots, {\bf{y}}_K \right]^{T} $, $ {\bf{H}}_s = \left[ {\bf{h}}_1, {\bf{h}}_2, \cdots, {\bf{h}}_K \right]^{T} \in \mathbb{C}^{K \times N_{\rm{BS}}} $ is frequency-spatial domain channel, and $ {\bf{N}} = \left[ {\bf{n}}_1, {\bf{n}}_2, \cdots, {\bf{n}}_K \right]^{T} $.
Moreover, by exploiting the angular-domain compressibility of massive MIMO channels, we define the frequency-angle domain channel ${\bf{H}}_a$ as \cite{KML}
\begin{equation}\label{angulr}
{\bf{H}}_a = {\bf{H}}_s {\bf{F}}.
\end{equation}
In (\ref{angulr}), ${\bf{F}} \in \mathbb{C}^{N_{\rm{BS}} \times N_{\rm{BS}}}$ is given by
\begin{equation}
{\bf{F}} = {\bf{F}}_{N_h}^T \otimes {\bf{F}}_{N_v},
\end{equation}
where ${\bf{F}}_{N_h} \  ({\bf{F}}_{N_v})$ is an $N_h \times N_h \ (N_v \times N_v)$ discrete Fourier transform (DFT) matrix.
Obversely, the $k$-th row of ${\bf{H}}_a$, i.e., $ {\bf{H}}_a \left( k,: \right) $ is a sparse row vector, $1 \le k \le K$.
Thus, equation (\ref{frequency spatial}) can be rewritten as
\begin{equation}\label{problem}
{\bf{Y}} = {\bf{H}}_{a} {\bf{F}}^{H} {\bf{X}} + {\bf{N}} = {\bf{H}}_{a} {\widetilde{\bf{X}}} + {\bf{N}},
\end{equation}
where ${\widetilde{\bf{X}}} = {\bf{F}}^{H} {\bf{X}}$.

The proposed DNN architecture for jointly designing the pilot signals and channel estimator
is depicted in Fig. \ref{fig2:NN}, which contains a dimensionality reduction network and a reconstruction network.
It is worth pointing out that the weights of the dimensionality reduction network and parameters of the reconstruction network are obtained simultaneously according to the Adam algorithm \cite{Adam}, so the pilot signals and the channel estimator are jointly optimized. Our motivation of using the proposed DNN architecture is to jointly mimic the linear compressibility process of high-dimensional signals and the corresponding non-linear signal reconstruction process in an end-to-end approach.

\subsubsection{Pilot Signals}
Specifically,
we stack the real and imaginary parts of the $k$-th subcarrier's channel $[{{\rm{Re}}{\{{\bf{h}}_k\}}}, {{\rm{{Im}}}{\{{\bf{h}}_k\}}}]^T$ as the input of DNN.
Then the input channel data is multiplied by the pilot signals ${\widetilde{\bf{X}}}$ to generate the low-dimensional measurements $[{{\rm{Re}}{\{{\bf{y}}_k\}}}, {{\rm{Im}}{\{{\bf{y}}_k\}}}]^T$, which can be rewritten as
\begin{equation}\label{model}
\left[ \begin{array}{c}
         {\rm{Re}}{\{{\bf{y}}_k^T\}} \\
         [1mm]
         {\rm{Im}}{\{{\bf{y}}_k^T\}}
       \end{array} \right] = \left[ \begin{array}{c}
                               {\rm{Re}}{\{{\bf{h}}_k^T\}} \\
                               [1mm]
                               {\rm{Im}}{\{{\bf{h}}_k^T\}}
                             \end{array} \right] {\widetilde{\bf{X}}} + \left[ \begin{array}{c}
                                                                                 {\rm{Re}}{\{{\bf{n}}_k^T\}} \\
                                                                                 [1mm]
                                                                                 {\rm{Im}}{\{{\bf{n}}_k^T\}}
                                                                               \end{array} \right].
\end{equation}
In this model, the real and imaginary parts of each measurement is a weighted linear combination of all the input channel values corresponding to real and imaginary parts. 
This compression mechanism can be modeled as a fully-connected layer without the biases values and the nonlinear activation function, which can be shown in the green portion of Fig. \ref{fig2:NN}.

\subsubsection{Channel Estimator}
The rest of the proposed DNN architecture is divided into two cascaded parts to reconstruct the high-dimensional channel from the low-dimensional measurements.
The first stage utilizes a fully-connected layer
to obtain an initial coarse channel estimation $[{{\rm{Re}}{\{\tilde{{\bf{h}}}_k\}}}, {{\rm{{Im}}}{\{\tilde{{\bf{h}}}_k\}}}]^T$, which can be used to imitate the matching operation (calculating the inner product between the measurement matrix and the measurements) of greedy CS algorithms, e.g., orthogonal matching pursuit (OMP) algorithm, to obtain an initial solution for starting the following iteration reconstruction operations.
In the second stage, to exploit the angular-domain compressibility of the massive MIMO channels, we first transform the dimensions of the coarse channel estimation $(2, {N_{\rm{BS}}})$ into $ (2, {N_h}, {N_v}) $.
Then, the coarse channel estimation is delivered to $N_{\rm{re}}$ refining units, and each of them contains a convolutional layer cascaded with a batch normalization (BN) layer and the leaky rectified linear unit (Leaky ReLU) activation function.
Specifically, in the $n$-th refining unit, for $ 1 \le n \le N_{\rm{re}} $, the outputs of the previous layer are processed by the convolutional layer with the kernel size of $3 \times 3$, and the number of convolutional filters is $2^n$. Zero padding (ZP) operation is added to maintain the dimension of the input channel unchanged after convolution, and the BN layer can avoid the gradient diffusion and overfitting.
It is worth mentioning that we also introduce a residual connection which directly transmits the data flow to the last layer. In this way, the vanishing gradient problem caused by multiple stacked non-linear transformations can be effectively avoided.

To define the difference between the input and output, the MSE over all the training samples has been employed, and its expression can be written as
\begin{equation}\label{lose:mse}
L({\bf{\Theta}}) = \frac{1}{P} \sum_{p=1}^{P} \| \widehat{\mathbf{h}}^{p} - {\mathbf{h}}^{p} \|_{F}^{2} = \frac{1}{P} \sum_{p=1}^{P} \| f ( {\mathbf{h}}^{p}, {\bf{\Theta}} ) - {\mathbf{h}}^{p} \|_{F}^{2},
\end{equation}
where $P$ is the number of the samples in each batch of the training set, ${\mathbf{h}}^{p}$ is the $p$-th channel sample in this batch, and the $p$-th estimated channel ${{\bf{\hat h}}^p} = f({{\bf{h}}^p},\bf{\Theta} )$ is a function of the input channel sample ${{\bf{h}}^p}$ and DNN's weights $\bf{\Theta}$.
Specifically, ${\bf{\Theta}} {\rm{ = }}\left\{ {{\bf{\tilde X}},{\bf{X'}},{{\bm{\theta}} _{{\rm{conv}}}}} \right\}$, where ${\bf{\tilde X}}$ is the weight of dimensionality reduction network, ${\bf{X'}}$ (weight of fully-connected network) and ${\bm{\theta}_{\rm{conv}}}$ (weight of CNN) are the weights of reconstruction network.
We jointly train $\bf{\Theta}$ in the end-to-end DNN architecture by executing ${{{\bf{\hat h}}}^p} = f({{\bf{h}}^p},{\bf{\Theta}} ) = {f^2}({f^1}({{\bf{h}}^p},{\bf{\tilde X}}),{\bf{X'}},{\bm{\theta} _{\rm{conv}}})$, where ${f^1\left( \right)}$ and ${f^2\left( \right)}$ denote the dimensionality reduction network and reconstruction network, respectively, and $\bf{\Theta}$ is optimized by using Adam algorithm with learning rate 0.001.
In addition, the batch size and epochs are set to 128 and 300, respectively.

\section{Simulation Results}
In this section, we investigate the performance of the proposed DNN-based channel estimation scheme. The simulation parameters are listed as follows: the BS is equipped with UPA with $N_h$ = $N_v$ = 16 so that $N_{\rm{BS}}$ = 256, the number of OFDM subcarriers is $K$ = 256, and the channel estimation compression ratio is $\rho$ = $M/N_{\rm{BS}}$.
In addition, the simultaneous orthogonal matching pursuit (SOMP)-based channel estimation scheme \cite{WC} is adopted for comparison, where the number of quantized angle grids $G$ is set to 64 for the dimensions of horizontal and vertical antennas, i.e., a redundant dictionary with oversampling ratio $G / N_h = G / N_v = 4$ is considered.
The experiments are performed in PyCharm Community Edition (Python 3.6 environment and Keras 2.2.4 whose backend is Tensorflow) on a computer with dual Intel Xeon 8280 CPU (2.6GHz) and dual Nvidia GeForce GTX 2080Ti GPUs.

For the proposed DNN structure, we generate training set including $S_{\rm{tr}}$ = 2,000 channel data groups according to certain channel scenarios, which will be explained in detail, and the dimension of input data in each group is $ \left( K, 2, N_{\rm{BS}} \right)$. We stack all the data along the first dimension to construct the overall data form \cite{noFD}, i.e, $\left( S_{\rm{tr}} \times K, 2, N_{\rm{BS}} \right)$. Similarly, the validation set and the test set are $S_{\rm{va}}$ = 1,000 groups and $S_{\rm{te}}$ = 500 groups, respectively. In other words, the training, validation, and testing sets contain 512,000, 256,000, 128,000 samples, respectively. The number of refining units in channel estimator is $N_{\rm{re}}$ = 8. We choose the normalized mean square error (NMSE) as the metric for performance evaluation, defined as
\begin{equation}\label{NMSEformula}
\mathrm{NMSE} = 10 \log_{10} \left( \mathbb{E} \left[ \| \mathbf{H} - \widehat{\mathbf{H}} \|_{F}^{2} / \| \mathbf{H} \|_{F}^{2} \right] \right).
\end{equation}

\subsection{Low-Frequency Massive MIMO over One-Ring Channel Model}
\begin{figure}[t]
\centering
{\subfigure[]{ \label{fig4:onering_NMSE(a)}
\centering
\includegraphics[scale=0.49]{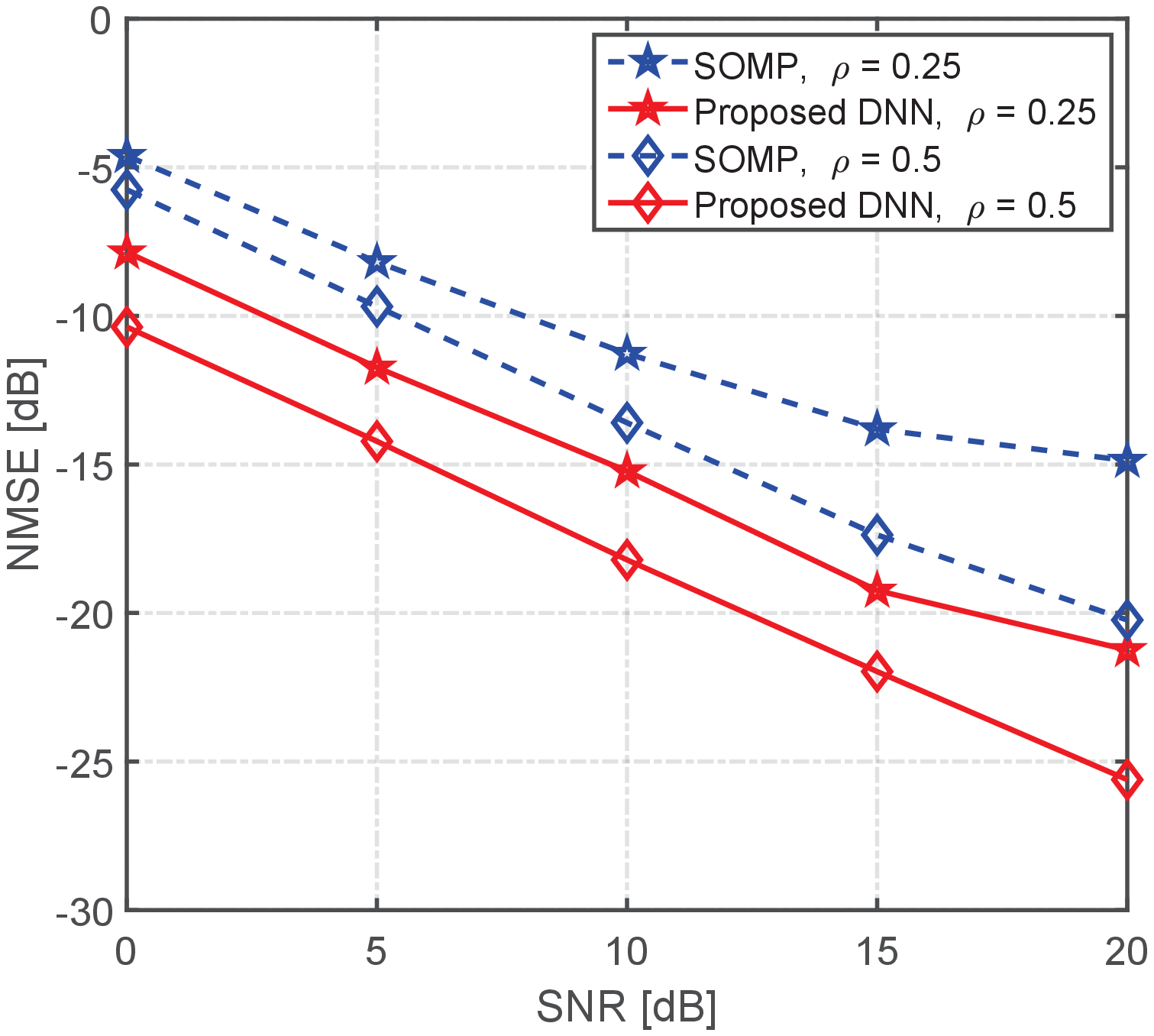}
}}
{\subfigure[]{ \label{fig4:onering_NMSE(b)}
\centering
\includegraphics[scale=0.49]{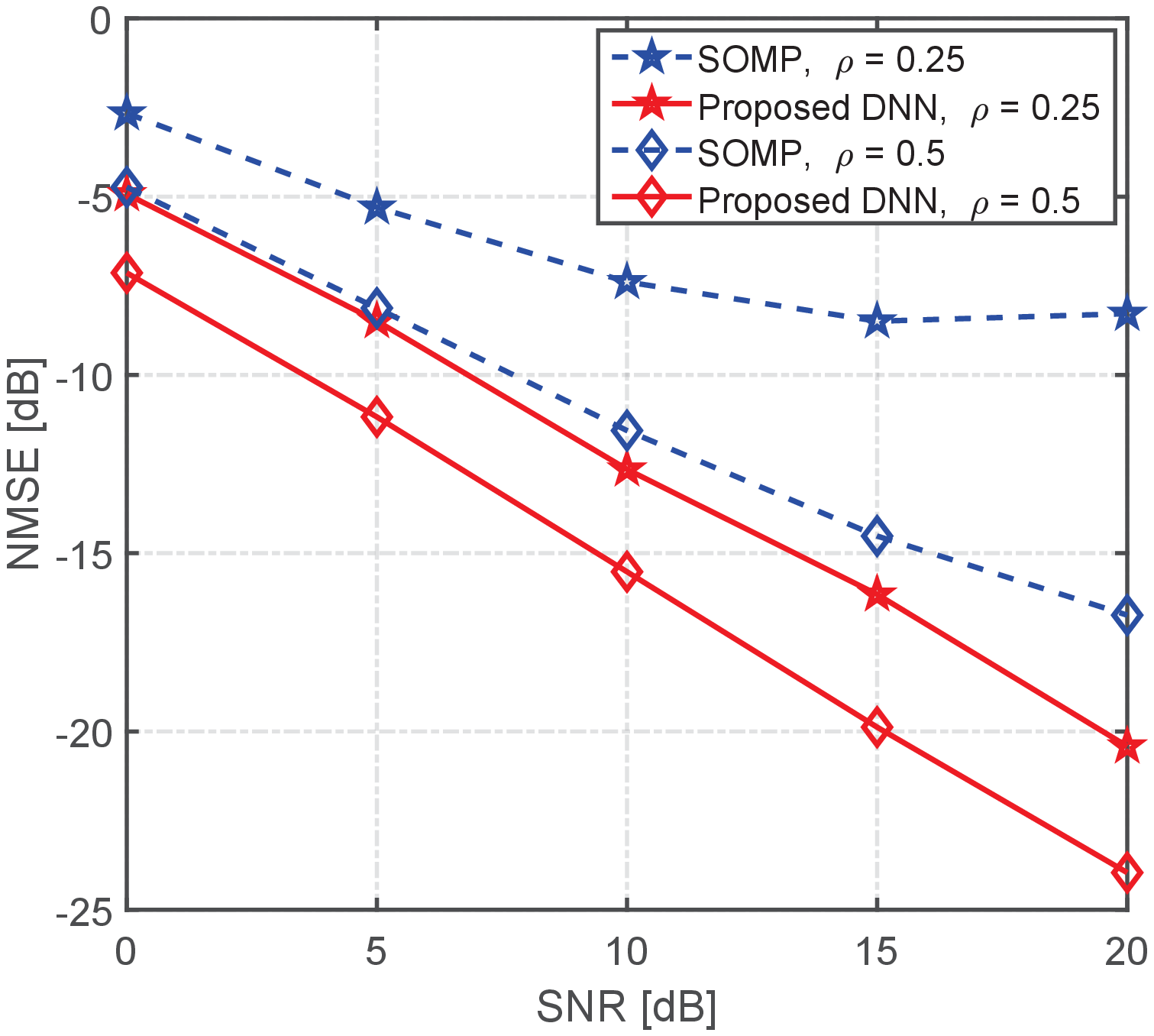}
}}
\caption{NMSE performance comparison of different channel estimation schemes for low-frequency massive MIMO systems based on one-ring channel model versus SNRs, where (a) angle spread $\Delta \theta = \pm 7.5^{\circ}$, (b) angle spread $\Delta \theta = \pm 15^{\circ}$.}
\label{fig4:onering_nmse}
\vspace{-5mm}
\end{figure}
For the one-ring channel model, we consider the number of paths is $N_c$ = 1, $N_p$ = 100, and the angle spreads $ \Delta \theta $ are $\pm 7.5^{\circ}$ and $\pm 15^{\circ}$, respectively.
Fig. \ref{fig4:onering_nmse} compares the NMSE performance of the proposed DNN-based channel estimation scheme with the SOMP-based channel estimation scheme against different signal-to-noise ratios (SNRs), where the different compression ratios $\rho$ and angle spreads $\Delta \theta$ are investigated.
We can observe that the proposed DNN-based channel estimation scheme with
$\rho$ = 0.25 even outperforms the SOMP-based channel estimation scheme with $\rho$ = 0.5.
This behavior indicates the proposed DNN-based scheme can acquire the better performance by using less pilot overhead.
It is because the DNN is capable of optimizing the weights and capturing the intrinsic characteristics of the channels by learning from extensive channel samples.
In this way, the accurate channel can be reconstructed with much reduced pilot overhead.
\subsection{MmWave Massive MIMO over Cluster Sparse Channel Model}
\begin{figure}[t]
\centering
{\subfigure[]{ \label{fig5:cluster_nmse}
\centering
\includegraphics[scale=0.49]{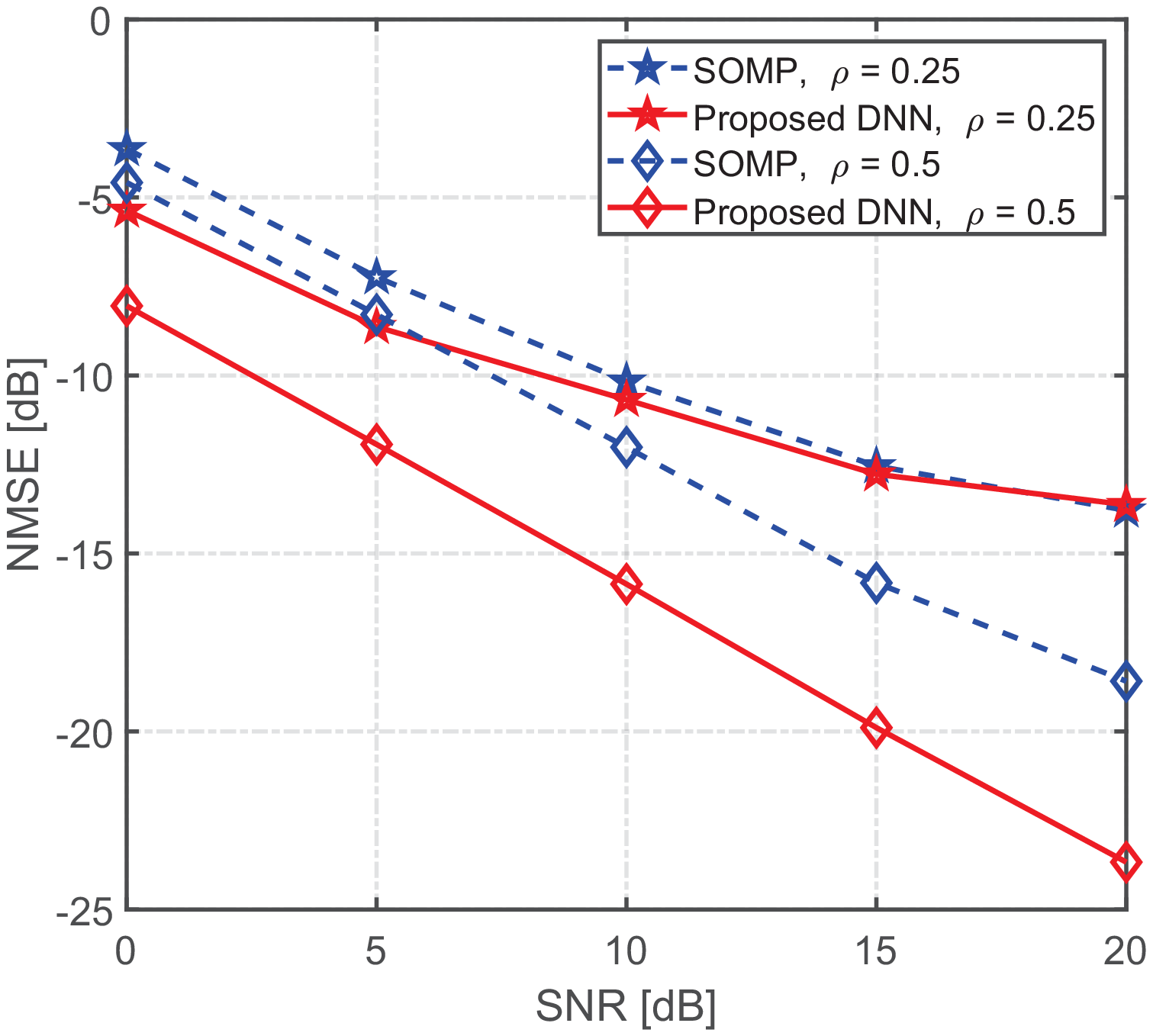}
}}
{\subfigure[]{ \label{fig6:subcarrier_nmse}
\centering
\includegraphics[scale=0.49]{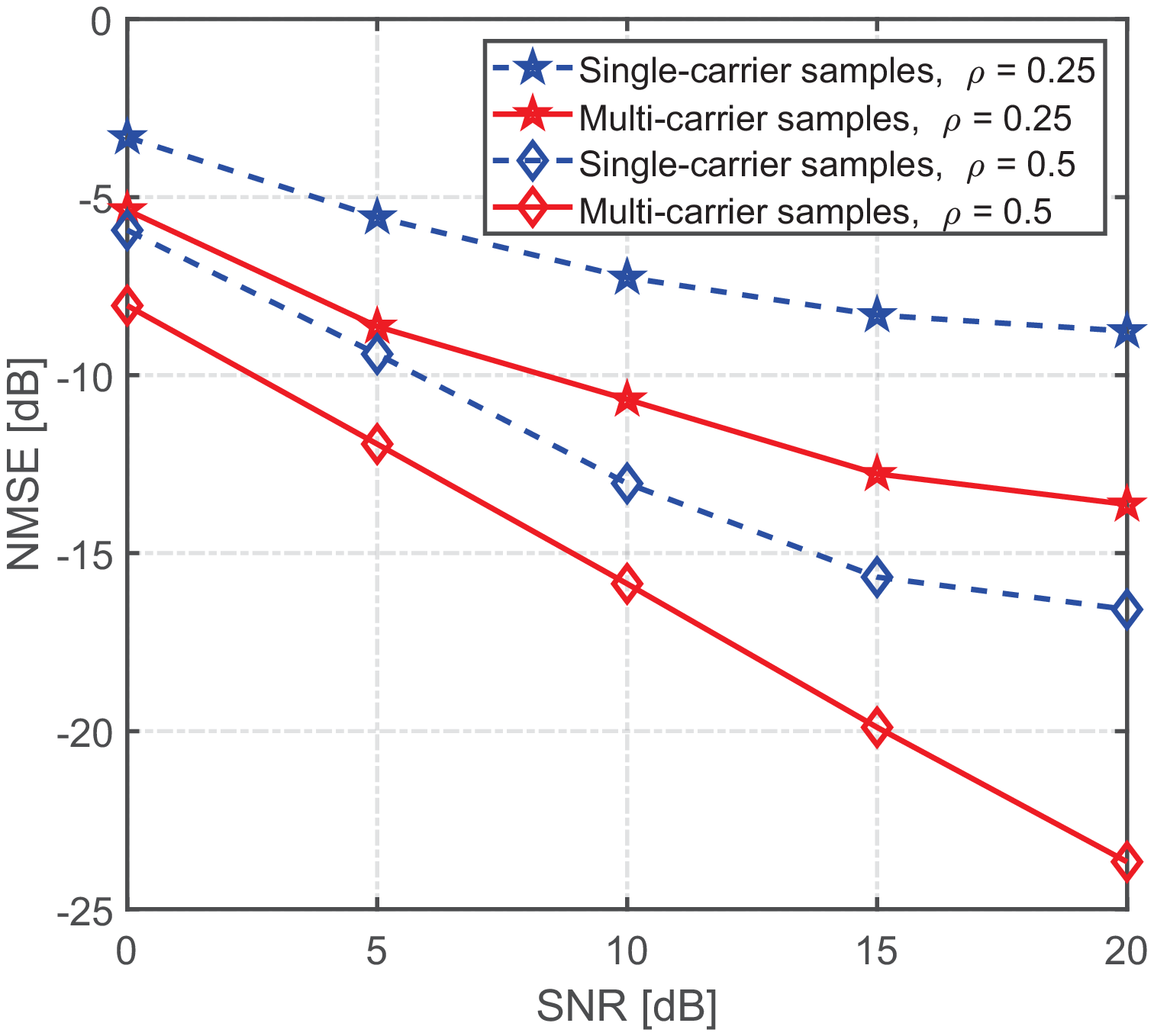}
}}
\caption{NMSE performance comparison of different channel estimation schemes for mmWave massive MIMO systems based on cluster sparse channel model versus SNRs, where (a) angle spread $\Delta \theta = \pm 3.75^{\circ}$, (b) the effectiveness of the proposed scheme in multi-subcarrier processing.}
\label{}
\vspace{-5mm}
\end{figure}

%
In the cluster sparse channel scenario, we consider $N_c$ = 6, $N_p$ = 10, and the angle spread $ \Delta \theta $ is $\pm 3.75^{\circ}$.
Fig. \ref{fig5:cluster_nmse} compares the NMSE performance of different channel estimation schemes versus SNRs with $\rho$ = 0.25, 0.5, respectively.
We can observe that the proposed DNN-based channel estimation scheme exceeds the SOMP-based channel estimation scheme, regardless of the compression ratio being 0.25 or 0.5.
Even in the low SNR regime, the proposed DL-based approach can achieve a better NMSE performance with less pilot overhead. It is worth pointing out that the slope of the DNN curve decreases as the SNR increases after 15 dB for the compression ratio $\rho$ = 0.25. We can find a reasonable explanation in \cite{slope} that the NMSE performance is not only affected by the SNR but also related to the number of training set samples and the setting of hyper-parameters.

Note that the effectiveness of the proposed DNN-based channel estimation scheme for OFDM systems is based on the multi-carrier channel samples, which consist of the channels of all subcarriers from different channel realizations.
To verify the effectiveness of multi-carrier channel samples, we also compare the performance of the proposed scheme based on single carrier channel samples, as shown in Fig. \ref{fig6:subcarrier_nmse}.
We can find the proposed DNN-based channel estimation scheme can exhibit more excellent performance by utilizing multi-subcarrier channel samples.

Finally, we compare the computation complexities of the proposed DNN-based channel estimation scheme and the SOMP-based channel estimation scheme.
Note that, in our proposed scheme, we train the neural network offline using the channel simulated values, which are generated according to the certain channel scenarios. In the case of offline learning, computational complexity is a less concern, because the required time is usually not strictly limited.
Once the proposed scheme has been trained, the obtained pilot signals and the channel estimator can be effectively used for an online channel estimation process.
The computation complexity of the proposed scheme in testing stage includes two fully-connected operations and multiple cascaded convolutional operations.
The computation complexity of the former is ${C_{\rm{fc}}} \sim {\mathcal{O}}(2N_{{\rm{BS}}}^2\rho )$ and that of the latter is ${C_{\rm{conv}}} \sim {\mathcal{O}}({N_h}{N_v}{k^2}\sum\limits_{l = 1}^{N_{\rm{re}}} {{n_{l - 1}}{n_l}} )$, where $k$ is the side length of the convolutional filters, $n_{l-1}$ and $n_{l}$ denote the numbers of input and output feature maps of the $l$th convolutional layer, $1 \le l \le {N_{{\rm{re}}}}$, respectively. Then the computational complexity of the proposed DNN-based channel estimation scheme is given in Table \ref{DNN_CC}.
Besides, the computational complexity of the SOMP-based channel estimation scheme is given in Table \ref{SOMP_CC} for comparison, where the number of iterations for the SOMP algorithm is denoted by ${\it{I}}$.
\begin{table}[ht]
  \renewcommand\arraystretch{2.5}
  \centering
  \caption{Computational Complexity of Proposed DNN-Based Scheme}\label{DNN_CC}
  \begin{tabular}{c|c}
    \Xhline{1.2pt}
    {\bf{Operation}} & {\bf{Complexity}} \\
    \Xhline{1.2pt}
    \makecell[c]{Fully-connected layer \\ in dimensionality reduction network} & ${\mathcal{O}} \left( N_{{\rm{BS}}}^2\rho \right)$ \\
    \hline
    \makecell[c]{Fully-connected layer \\ in reconstruction network} & ${\mathcal{O}} \left( N_{{\rm{BS}}}^2\rho \right)$ \\
    \hline
    \makecell[c]{Convolutional layers \\ in reconstruction network} & ${\mathcal{O}} \left( {N_h}{N_v}{k^2}\sum\limits_{l = 1}^{N_{\rm{re}}} {{n_{l - 1}}{n_l}} \right)$ \\
    \Xhline{1.2pt}
  \end{tabular}
\end{table} \\
\begin{table}[ht]
\vspace{-4mm}
  \renewcommand\arraystretch{2.0}
  \centering
  \caption{Computational Complexity of SOMP-Based Scheme}\label{SOMP_CC}
  \begin{tabular}{c|c}
    \Xhline{1.2pt}
    {\bf{Operation}} & {\bf{Complexity}} \\
    \Xhline{1.2pt}
    Correlation & ${\mathcal{O}} \left( {\rho {N_{{\rm{BS}}}}{G^2}KI} \right)$ \\
    \hline
    Project subspace & \makecell[c]{${\mathcal{O}} \left( {\frac{1}{4}{I^2}{{(I + 1)}^2} + \frac{1}{3}\rho {N_{{\rm{BS}}}}I(I + 1)(2I + 1)} \right.$ \\ $\left. { + \frac{1}{2}\rho {N_{{\rm{BS}}}}KI(I + 1)} \right)$} \\
    \hline
    Update residual & ${\mathcal{O}} \left( {\frac{1}{2}\rho {N_{{\rm{BS}}}}KI(I + 1)} \right)$ \\
    \hline
    Compute MSE & ${\mathcal{O}} \left( {\rho {N_{{\rm{BS}}}}{K^2}I} \right)$ \\
    \hline
    Reestablishment & ${\mathcal{O}} \left( {{N_{{\rm{BS}}}}KI} \right)$ \\
    \Xhline{1.2pt}
  \end{tabular}
\end{table}
\vspace{-5mm}
\section{Conclusions}
In this paper, we have proposed a DNN-based channel estimation scheme to jointly design the pilot signals and channel estimator.
Specifically, we model the pilot-based channel training and channel reconstruction as an end-to-end DNN, which consists of a dimensionality reduction network and a channel reconstruction network.
For this DNN, we regard the input and weights of the fully-connected layer in dimensionality reduction network as the channel and pilot signals, respectively.
From the low-dimensional measurements, the high-dimensional channel will be estimated from the following reconstruction network, which is composed of a fully-connected layer and multiple cascaded convolutional layers.
Simulation results have verified that the proposed DNN-based channel estimation scheme can achieve significant improvement in channel estimation performance than the conventional scheme in both the low-frequency one-ring channel model and mmWave frequency with cluster sparse channel model.

\end{document}